\documentclass{lor-proc}

\usepackage{graphics,epsfig}
\usepackage{psfrag}

\newtheorem{theorem}{Theorem}[section]

\theoremstyle{definition}
\newtheorem{definition}[theorem]{Definition}

\theoremstyle{remark}

\numberwithin{equation}{section}

\copyrightinfo{}
{}

\begin{document}

\bibliographystyle{amsalpha}

\title{On timelike surfaces in Lorentzian manifolds}

\author{Wolfgang Hasse}
\address{Wolfgang Hasse, Institute of Theoretical Physics, 
TU Berlin, Hardenbergstr.36, 
10623 Berlin, Germany, and Wilhelm Foerster Observatory Berlin, Munsterdamm 90, 12169 Berlin, Germany}
\email{astrometrie@gmx.de}

\author{Volker Perlick}
\address{Volker Perlick, Physics Department, Lancaster University, 
Lancaster LA1 4YB, United Kingdom}
\email{v.perlick@lancaster.ac.uk}

\date{April 2008}



\begin{abstract}
We discuss the geometry of timelike surfaces (two-dimensional
submanifolds) in a Lorentzian manifold and its interpretation 
in terms of general relativity. A classification of such surfaces 
is presented which distinguishes four cases of special algebraic 
properties of the second fundamental form from the generic case. 
In the physical interpretation a timelike surface $\Sigma$ can 
be viewed as the worldsheet of a ``track'', and timelike curves 
in $\Sigma$ can be viewed as the worldlines of observers who are 
bound to the track, like someone sitting in a roller-coaster car. 
With this interpretation, our classification turns out to be 
closely related to (i) the visual appearance of 
the track, (ii) gyroscopic transport along the track, and 
(iii) inertial forces perpendicular to the track. We illustrate
our general results with timelike surfaces in the Kerr-Newman 
spacetime.

\end{abstract}

\maketitle

\section{Introduction}\label{sec:intro}

A standard tool for investigating the geometry of a submanifold in
a semi-Riemannian manifold is the second fundamental
form, or shape tensor, see e.g. O'Neill \cite{ONeill1983}). In this
paper we will discuss some aspects of the second fundamental form 
for the case that the ambient semi-Riemannian 
manifold has Lorentzian signature $(-,+, \dots, + )$ and that the 
submanifold is two-dimensional with Lorentzian signature $(-,+)$. 
We call such a submanifold a \emph{timelike surface} for short. 
Timelike surfaces are interesting objects not only from a mathematical
point of view but also in view of physics. A four-dimensional Lorentzian
manifold can be interpreted as a spacetime in the sense of general
relativity, and a timelike surface $\Sigma$ in such a manifold can be 
interpreted as the worldsheet of an object with one spatial dimension.
It is often helpful to think of $\Sigma$ as being realized by a ``track'',
and of timelike curves in $\Sigma$ as being the worldlines of observers who
are bound to the track like someone sitting in a roller-coaster car, cf.
Abramowicz \cite{Abramowicz1990}.

It is the main goal of this paper to give a classification of timelike
surfaces in terms of their second fundamental form, and to discuss the
physical relevance of this classification in view of the roller-coaster
interpretation. As we want to concentrate on properties of timelike
surfaces which are conformally invariant, we base our classification 
on the trace-free part of the second fundamental form. We call a timelike
surface ``generic'' if this trace-free part is non-degenerate, in a 
sense specified below, and ``special'' otherwise. It turns out that a
degeneracy can occur in four different ways, giving rise
to four different types of special timelike surfaces. Using the 
roller-coaster interpreation, we will see how
each of the four special types can be distinguished from generic 
timelike surfaces by three observational features: (i) the visual 
appearance of the track, (ii) gyroscopic transport along the track, 
and (iii) inertial forces perpendicular to the track. For the latter
we use the definition of inertial forces given in 
\cite{FoertschHassePerlick2003}. It can be viewed as an 
adaptation to general relativity of Huygens' definition of 
centrifugal force, which was based 
on the curvature of a track. For the history of this notion see 
Abramowicz \cite{Abramowicz1990}.

Our discussion of the physics of timelike surfaces is kinematic, as
opposed to dynamic, in the sense that Einstein's field equation is 
not used and that we do not specify an equation of motion for our 
timelike surfaces. As outlined above, our main physical motivation
is to give an operational approach to inertial forces and its relation
with the visual appearance of a track and with gyroscopic transport.
However, we would like to mention that our results might also be of
interest in view of applications to strings. The worldsheet of  a 
(classical) string is a timelike surface, and the second fundamental 
form of this surface gives some information on the physical properties
of the string, see e.g. Carter \cite{Carter1995}. 

The paper is organized as follows. In Section \ref{sec:timesur} 
we introduce orthonormal basis vector fields and 
lightlike basis vector fields on timelike surfaces and we 
classify such surfaces in terms of their second fundamental form. 
The physical interpretation of timelike surfaces, based on the
roller-coaster picture, and its relation to the second fundamental 
form is discussed in Section \ref{sec:interpretation}. Among
other things, in this section we consider two quite different 
splittings of the total inertial force perpendicular to the track 
into three terms and we discuss the invariance properties of 
these terms. Generic timelike surfaces are treated in Section 
\ref{sec:generic}; we show that the relevant properties of 
the second fundamental form are encoded in a characteristic 
hyperbola and that every generic timelike surface admits a 
distinguished reference frame (timelike vector field). The next four 
short sections are devoted to the four non-generic cases and to the
observable features by which each of them differs from the generic
case. Finally, in Section \ref{sec:example} we illustrate our 
results with timelike surfaces in the Kerr-Newman spacetime.

\section{Timelike surfaces}\label{sec:timesur}
Let $(M,g)$ be an $n$-dimensional Lorentzian manifold. We assume
that the metric $g$ is of class $C^{\infty}$ and we choose the 
signature of $g$ to be $(-, + , \dots, +)$. It is our goal to 
investigate the geometry of surfaces (i.e., two-dimensional 
$C^{\infty}$-submanifolds) of $M$ that are timelike everywhere,
i.e., the metric pulled back to the surface is supposed to have
signature $(-,+)$. If we fix such a surface $\Sigma$, we may choose
at each point $p \in \Sigma$ an orthonormal basis for the tangent 
space $T_p \Sigma$. We assume that this can be done globally on
$\Sigma$ with the basis depending smoothly on the foot-point.
This means that we assume $\Sigma$ to be time-orientable and 
space-orientable. This gives us two $C^{\infty}$ vector fields
$n$ and $\tau$ on $\Sigma$ that satisfy
\begin{equation}\label{eqn-defntau}
g(\tau, \tau ) = - g(n,n) = 1 \; , \qquad g(n, \tau ) = 0 \, .
\end{equation}

At each point, $n$ and $\tau$ are unique up to a two-dimensional
Lorentz transformation $(n, \tau ) \longmapsto ( n' \,\tau ')$.
If we restrict to transformations that preserve the time-orientation,
i.e., if we require that the two timelike vectors $n$ and $n'$ 
point into the same half of the light cone, any such Lorentz
transformation is of the form
\begin{equation}\label{eq:trafontau}
n' = \frac{n+v \tau}{\sqrt{1-v^2}} \; , \qquad
\tau ' = \frac{v n + \tau}{\sqrt{1-v^2}} 
\end{equation}
where the number $v$ gives the velocity of the $n'$-observers 
relative to the $n$-observers, in units of the velocity of 
light, $-1 < v < 1$. Of course, $v$ may vary from
point to point. 

From the orthonormal basis $(n , \tau )$ we may switch to
a lightlike basis $(l_+,l_-)$ via 
\begin{equation}\label{eq:defl}
l_{\pm} = n \pm \tau \; .
\end{equation}
Under a Lorentz transformation (\ref{eq:trafontau}) this
lightlike basis transforms according to
\begin{equation}\label{eq:trafol}
l _{\pm} ' = \frac{1 \pm v}{\sqrt{1-v^2}} \, l _{\pm} \; .
\end{equation}
Thus, the directions of $l_+$ and $l_-$ are invariant with 
respect to Lorentz transformations. This reflects the obvious 
fact that, at each point of the timelike surface $\Sigma$, 
there are precisely two lightlike directions tangent to $\Sigma$.
The integral curves of $l_+$ and $l_-$ give two families
of lightlike curves each of which rules the surface $\Sigma$.
We want to call $\Sigma$ a \emph{photon surface} if both families
are geodesic, and we want to call $\Sigma$ a \emph{one-way photon 
surface} if one of the two families is geodesic but the other 
is not. Clearly, this terminology refers to the fact that in 
general relativity a lightlike geodesic is interpreted as
the worldline of a (classical) photon. More generally, one can define
a photon $k$-surface to be a $k$-dimensional submanifold 
of a Lorentzian manifold for which every lightlike geodesic
that starts tangent to $\Sigma$ remains tangent to $\Sigma$.
This notion was introduced, for the case $k = n-1$, in a paper
by Claudel, Virbhadra and Ellis \cite{ClaudelVirbhadraEllis2001};
here we are interested in the case $k=2$ which was already 
treated in \cite{FoertschHassePerlick2003} and 
\cite{Perlick2005}. 

Before discussing photon surfaces and one-way photon surfaces, 
we will demonstrate that these notions appear
naturally when timelike surfaces are classified with respect
to their second fundamental form. To work this out, we recall
(cf. e.g. O'Neill \cite{ONeill1983}) that the \emph{second fundamental 
form}, or \emph{shape tensor field}, $\Pi$ is well-defined for any 
nowhere lightlike submanifold of a semi-Riemannian manifold, in 
particular for a timelike surface $\Sigma$ of a Lorentzian manifold, 
by the equation
\begin{equation}\label{eqn-defPi}
\Pi (u, w) \, = \, P^{\perp} \big( \nabla _u w \big) \, 
\end{equation}
where $u$ and $w$ are vector fields on $\Sigma$. Here $\nabla$ is
the Levi-Civita connection of the metric $g$ and $P^{\perp}$ denotes the 
orthogonal projection onto the orthocomplement of $\Sigma$,
\begin{equation}\label{eq:defP}
P^{\perp}(Y) = Y - g(\tau,Y) \, \tau + g(n,Y) \, n \, .
\end{equation}
As $\nabla _u w - \nabla _w u = [u,w]$ must be tangent to $\Sigma$, 
we can read from (\ref{eqn-defPi}) the well-known fact that $\Pi$ 
is a symmetric tensor field along $\Sigma$.

With respect to the lightlike basis $(l_+,l_-)$, the second fundamental
form is characterized by its three components 
\begin{equation}\label{eq:defPipm}
\Pi _+ = \Pi (l_+,l_+) \, , \qquad \Pi _- = \Pi (l_-,l_-) \, ,
\qquad \Pi _0 = \Pi (l_+, l_-) = \Pi ( l_-,l_+) \, .
\end{equation}
Note that these three vectors lie in the orthocomplement of $\Sigma$,
so they are necessarily spacelike. In a 4-dimensional Lorentzian 
manifold, this orthocomplement is two-dimensional, so $\Pi_+, \Pi_-$ 
and $\Pi_0$ must be linearly dependent. In higher-dimensional 
Lorentzian manifolds, however, these three vectors may be 
linearly independent.

From (\ref{eq:trafol}) we can read the transformation behaviour of
$\Pi _+, \Pi_-$ and $\Pi _0$ under Lorentz transformations,
\begin{equation}\label{eq:trafoPi}
\Pi _{\pm} ' = \frac{1 \pm v}{1 \mp v} \, \Pi _{\pm} \; , \qquad
\Pi _0 ' = \Pi _0 \, .
\end{equation}
Thus, $\Pi _+$ and $\Pi _-$ are invariant up to multiplication
with a positive factor whereas $\Pi _0$ is invariant. In particular,
the conditions $\Pi _+ =0$ and $\Pi _- =0$ have an invariant meaning.
Similarly, the statement that $\Pi _+$ and $\Pi _-$ are parallel
(or anti-parallel, respectively) has an invariant meaning.

Note that 
$\Pi _0$ is related to the trace of the second fundamental form by
\begin{equation}\label{eq:tracePi}
\Pi _0 = -  \text{trace} ( \Pi ) \, .
\end{equation}

We now introduce the following terminology.

\begin{definition}\label{def:generic}
A timelike surface $\Sigma$ is called \emph{generic} (at $p$) if
$\Pi _+$ and $\Pi _-$ are linearly independent (at $p$). Otherwise 
it is called \emph{special} (at $p$).
\end{definition}

Clearly, the class of special timelike surfaces can
be subdivided into four subclasses, according to the 
following definition.

\begin{definition}\label{def:specialsub}
A timelike surface $\Sigma$ is called 
\begin{itemize}
\item[(a)]
\emph{special of the first kind} (at $p$) if $\Pi _+$ and
$\Pi _- $ are both non-zero and parallel, $\Pi _- = \alpha
\Pi _+$ with $\alpha >0$ (at $p$);
\item[(b)]
\emph{special of the second kind} (at $p$) if $\Pi _+$ and
$\Pi _- $ are both non-zero and anti-parallel, $\Pi _- = \alpha
\Pi _+$ with $\alpha <0$ (at $p$);
\item[(c)]
\emph{special of the third kind} (at $p$) if one of the 
vectors $\Pi _+$ and $\Pi _- $ is zero and the other is non-zero
(at $p$);
\item[(d)]
\emph{special of the fourth kind} (at $p$) if both $\Pi _+$ and 
$\Pi _- $ are zero (at $p$).
\end{itemize}
\end{definition}

Photon surfaces are timelike surfaces that are special of the fourth 
kind, whereas one-way photon surfaces are timelike surfaces that are
special of the third kind.  

It is obvious from (\ref{eq:trafoPi}) that the property of being generic 
or special of the $N$th kind is independent of the chosen orthonormal
frame. Moreover, it is preserved under conformal transformations. If we
multiply the metric $g$ with a conformal factor $e^{2f}$, where
$f$ is a function on $M$, and rescale the basis vectors accordingly, 
\begin{equation}\label{eq:conformal} 
\tilde{g} = e^{2f} g \, , \quad
\tilde{n} = e^{-f} n \, , \quad
\tilde{\tau} = e^{-f} \tau \, ,
\end{equation}
$\Pi _+$ and $\Pi_-$ are unchanged, whereas $\Pi _0$ transforms 
inhomogeneously,
\begin{equation}\label{eq:conformalPi} 
\tilde{\Pi} {}_+ = \Pi _+ \, , \quad
\tilde{\Pi} {}_- = \Pi _- \, , \quad
\tilde{\Pi} {}_0 = \Pi _0 + 2 P^{\perp} (U) \, , 
\end{equation}
where $df = g(U, \, \cdot \, )$. Thus, it is always possible to make
$\Pi _0$ equal to zero by a conformal transformation. This is
the reason why we based our classification on $\Pi _+$ and $\Pi _-$
alone. 

We will now review the physical interpretation connected with
$\Pi _+$ and $\Pi _-$, and then discuss the different types
of timelike surfaces one by one.

\section{Physical interpretation}\label{sec:interpretation}
If $M$ is 4-dimensional, $(M,g)$ can be interpreted as a spacetime
in the sense of general relativity. As indicated already in the 
introduction, we may interpret each timelike surface $\Sigma$ 
as the worldsheet of a track and each timelike curve in $\Sigma$ 
as the worldline of an observer who sits in a roller-coaster car
that is bound to the track. We want to 
discuss three types of ``experiments'' such an observer can carry
out, viz. (i) sending and receiving light rays, (ii) measuring the 
precession of gyroscopes, and (iii) measuring inertial accelerations. 
All three types of experiments turn out to be closely
related to the second fundamental form.

In general relativity light rays (i.e. worldlines of classical 
photons) are to be identified with lightlike geodesics. If an 
observer at one point of the track receives a light ray from
an observer at some other point of the track, the corresponding
lightlike geodesic will, in general, not arrive tangentially
to the track. Thus, the observer who receives the light ray
will get the visual impression that the track is curved. 
Photon surfaces are characterized by the property that such
a light ray always arrives tangentially to the track, i.e.,
a photon surface is the worldsheet of a track that appears
straight. In the case of a one-way photon surface this is true
only when looking in one direction (``forward''), but 
not when looking in the other direction (``backward''). 

If we have chosen an orthonormal basis $(n, \tau )$ on $\Sigma$,
we can interpret the integral curves of $n$ as observers 
distributed along the track described by $\Sigma$. If we 
want to give an interpretation  to $\tau$, we may think of
each of these observers holding a rod in the direction of
the track. We want to investigate whether $\tau$ can be 
realized as the axis of a gyroscope that is free to 
follow its inertia. According to general relativity, this
is true if and only if $\tau$ remains Fermi-Walker parallel
to itself along each integral curve of $n$ (see e. g. Misner, 
Thorne and Wheeler \cite{MisnerThorneWheeler1973}, 
Sect. 40.7 ), i.e., if and only if $\nabla _n \tau$ is a 
linear combination of $n$ and $\tau$. This is true if 
and only if $\Pi (n, \tau ) = 0$, which can be rewritten,
in terms of the lightlike vector fields (\ref{eq:defl}),
as $\Pi ( l_+ + l_- , l_+ - l_- ) = 0$. Using the 
notation from (\ref{eq:defPipm}), we find that
\begin{equation}\label{eq:gyro}
\Pi _+ = \Pi _-
\end{equation}
is the necessary and sufficient condition for $\tau$ 
being Fermi-Walker parallel along $n$. If (\ref{eq:gyro})
holds along an integral curve of $n$, a gyroscope carried
by the respective observer will remain parallel to the
track if it is so initially. Note that (\ref{eq:gyro})
is preserved under Lorentz transformations
(\ref{eq:trafoPi}) if and only if $\Pi _+ = \Pi _- = 0$.

Having chosen an orthonormal basis $(n , \tau )$ on $\Sigma$,
we can write any timelike curve in $\Sigma$ as the integral
curve of a vector field $n'$ that is related to $n$ by a 
Lorentz transformation according to (\ref{eq:trafontau}),
with a relative velocity $v$ that depends on the foot-point.
We want to calculate the vector  
$
\Pi ( n' , n' ) = P^{\perp} \big( \nabla _{n'} n' ) \; .
$
Using (\ref{eq:defl}), (\ref{eq:trafontau}) and 
(\ref{eq:defPipm}), we find
\begin{equation}\label{eq:Pinnv}
\Pi (n',n') = \frac{1}{2} \, \Pi _0 \, + \, 
\frac{1}{4} \, \frac{1+v}{1-v} \, \Pi _+ \, + \, 
\frac{1}{4} \, \frac{1-v}{1+v} \, \Pi _- \, .
\end{equation}
According to general relativity, the vector $\nabla _{n'} n'$ 
gives the acceleration of an observer traveling on an integral
curve of $n'$, measured relatively to a freely falling object.
If we think of this observer as sitting in a roller-coaster car
bound to the track modeled by $\Sigma$, the vector $- \Pi (n' n')$ 
gives the acceleration perpendicular to the track of a freely 
falling particle relative to the car. This relative acceleration 
is what an observer on a roller-coaster feels in his or her 
stomach, because the stomach wants to follow its inertia 
and move in free fall, whereas the frame of the observer's
body cannot follow this motion as it is strapped to the car.
For this reason, $- \Pi (n',n')$ is to be interpreted as the
(relativistic) \emph{inertial acceleration} of the $n'$-observers.
Multiplication with the mass gives the (relativistic) \emph{inertial
force} onto these observer. Following \cite{FoertschHassePerlick2003}, 
we can decompose the vector $- \Pi (n',n')$ into gravitational 
acceleration $a _{\text{grav}}$, Coriolis acceleration $a_{\text{Cor}}$ 
and centrifugal acceleration $a _{\text{cent}}$ by rearranging the 
right-hand side of (\ref{eq:Pinnv}) according to the following rule. 
$a _{\text{grav}}$ comprises all terms which are independent of 
$v$, $a_{\text{Cor}}$ comprises all terms of odd powers of $v$, 
and $a _{\text{cent}}$ comprises all terms of even powers of $v$,
\begin{equation}\label{eq:defgCc}
\Pi (n',n') =
\underset{- a_{\text{grav}}}{\underbrace{
\frac{1}{2} \Pi _0 + \frac{1}{4}
( \Pi _+ + \Pi _-)}}
+
\underset{-a_{\text{Cor}}}{\underbrace{
\frac{v}{1-v^2} \frac{1}{2} ( \Pi_+ - \Pi _- )}}
+
\underset{- a _{\text{cent}}}{\underbrace{
\frac{v^2}{1-v^2} \frac{1}{2} ( \Pi _+ + \Pi _- )}}
\; .
\end{equation}
(In \cite{FoertschHassePerlick2003} we found it convenient to work
with the corresponding covectors $A_{\text{grav}} = g \big ( 
a_{\text{grav}} , \, \cdot \, \big)$, etc.) This definition of 
gravitational, Coriolis and centrifugal accelerations with
respect to a timelike surface has the advantage that it is
unambiguous in an arbitrary Lorentzian manifold and that it
corresponds, as closely as possible, with the traditional
non-relativistic notions. 
(For alternative definitions of inertial accelerations in 
arbitrary general-relativistic spacetimes see, e.g. 
Abramowicz, Nurowski and Wex \cite{AbramowiczNurowskiWex1993}
or Jonsson \cite{Jonsson2006}.)

Whereas the decomposition (\ref{eq:defgCc}) depends on $n$, the 
splitting (\ref{eq:Pinnv}) of the inertial acceleration into three 
terms is invariant under Lorentz transformations. This follows from 
the transformation properties (\ref{eq:trafoPi}). For that and for 
some other calculations in this paper it is convenient to substitute 
\begin{equation}\label{eq:eta}
\frac{1}{\sqrt{1-v^2}} = \cosh \eta 
\quad \text{and} \quad
\frac{v}{\sqrt{1-v^2}} 
= \sinh \eta \, .
\end{equation}
Then the first equation in (\ref{eq:trafoPi}) reads 
$\Pi _{\pm} ' = e^{\pm 2 \eta} \, \Pi _{\pm}$. Similarly to the decomposition 
(\ref{eq:defgCc}) one can, owing to the different dependencies on 
$v$, operationally separate the three terms of the sum in 
(\ref{eq:Pinnv}) by measuring the inertial acceleration for different 
velocities.

\section{Generic timelike surfaces}\label{sec:generic}
In this section we consider a timelike surface which is generic at all points.
We can then characterize the second fundamental form at each point by the 
three non-zero vectors 
\begin{equation}\label{eq: Piinv}
\Pi _0 \, , \qquad I_+ = \sqrt{g(\Pi _- , \Pi _- )} \, \Pi _+ \, , 
\qquad I_- = \sqrt{g(\Pi_+, \Pi _+ )} \, \Pi _- 
\end{equation}
which, according to (\ref{eq:trafoPi}), are invariant with respect
to Lorentz transformations. Moreover, the two linearly independent
vectors $I _+$ and $I _-$ are conformally invariant, see
(\ref{eq:conformalPi}).

\begin{figure}[h]
    \psfrag{O}{O} 
    \psfrag{P}{\hspace{-0.4cm} $\tfrac{1}{2} \Pi _0$} 
    \psfrag{Q}{\hspace{-0.4cm} $\Pi _+$} 
    \psfrag{R}{\hspace{-0.5cm} $\Pi _-$} 
    \psfrag{1}{1} 
    \psfrag{2}{2} 
\centerline{\epsfig{figure=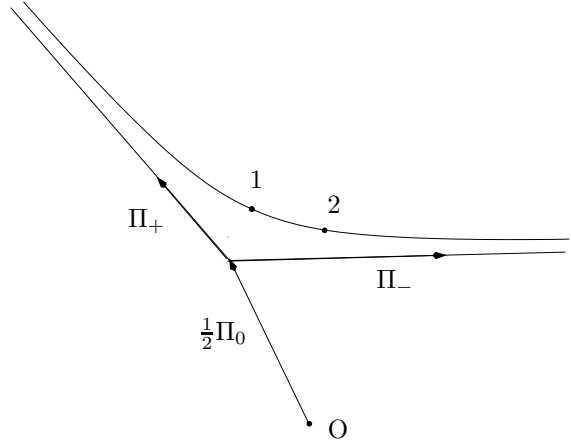,width=6.4cm,angle=15}}
\vspace{-1cm}
\caption{Characteristic hyperbola of the second 
fundamental form at a point $p$ of $\Sigma$.}\label{fig:hyper}
\end{figure}

We now fix an orthonormal basis $(n, \tau)$. Then we find all future-oriented
vector fields $n'$ with $g(n',n')=-1$ by a Lorentz transformation 
(\ref{eq:trafontau}). If $v$ runs from $-1$ to 1, at each point $p \in \Sigma$
the vector $\Pi (n',n')$ runs through a hyperbola, according to 
(\ref{eq:Pinnv}), see Figure \ref{fig:hyper}. We call this the 
\emph{characteristic hyperbola} of the second fundamental form at $p$.
The characteristic hyperbola lies in the orthocomplement $P^{\perp} (T_p M)$
of the tangent space $T_p \Sigma$, which is an $(n-2)-$dimensional
Euclidean vector space.

The asymptotes of the characteristic hyperbola are spanned by the 
linearly independent vectors $\Pi _+$ and $\Pi _-$ (or, what is the 
same, by the invariant vectors $I_+$ and $I_-$). The characteristic 
hyperbola is invariant with respect to
Lorentz transformations, see (\ref{eq:trafoPi}), whereas a conformal 
transformation produces a translation of the characteristic hyperbola, 
see (\ref{eq:conformalPi}).

The points on the characteristic hyperbola are in one-to-one correspondence
with future-oriented vectors normalized to $-1$ at $p$. Clearly, the
vertex of the hyperbola, indicated by 1 in Figure \ref{fig:hyper}, defines
a \emph{distinguished observer field} on every generic timelike surface. 
From (\ref{eq:Pinnv}) we find that the arrow from the origin to the vertex of the 
hyperbola is given by the vector
\begin{equation}\label{eq:turning}
\frac{1}{2} \, \Pi _0 \, + \, 
\frac{1}{4} \;
\sqrt[4]{\frac{g(\Pi _- , \Pi _- )}{g( \Pi _+ , \Pi _+ )}}
\; \Pi _+ 
\, + \, 
\frac{1}{4} \;
\sqrt[4]{\frac{g(\Pi _+ , \Pi _+ )}{g( \Pi _- , \Pi _- )}}
\; \Pi _- \; 
\end{equation}
which is Lorentz invariant, by (\ref{eq:trafoPi}). 
If we choose the distinguished observer field for our $n$, we have in the 
orthonormal basis $(n, \tau)$
\begin{equation}\label{eq:defdis}
g(\Pi _+ , \Pi _+ ) = g ( \Pi _- , \Pi _- ) \; .
\end{equation}
This property characterizes the distinguished observer field uniquely.

The distinguished observer field can be determined as the solution of 
an eigenvalue problem. Using the invariant vectors $I_{\pm}$ from 
(\ref{eq: Piinv}), we can introduce the real-valued bilinear form
\begin{equation}\label{eq:defpi}
\pi (u,w) \, = \, \frac{1}{2} \, g \big(  I_+ + I_- , \Pi (u,w) \big) 
\end{equation}
where $u$ and $v$ are tangent to $\Sigma$.
If $n$ is the distinguished observer field, the 
basis vectors $n$ and $\tau$ satisfy the eigenvalue equations
\begin{equation}\label{eq:eigen}
\pi (n , \, \cdot \, ) = \lambda _1 \, g( n, \, \cdot \, ) \; , \qquad
\pi (\tau , \, \cdot \, ) = \lambda _2 \, g( \tau, \, \cdot \, ) \; .
\end{equation}
To prove this we observe that, for an arbitrary orthonormal frame
$(n, \tau)$, 
\begin{equation}\label{eq:proofpi}
\pi ( n , \tau ) \, = \, \frac{1}{8} \,
\big( \, \sqrt{g(\Pi _- , \Pi _- )} - \sqrt{g(\Pi _+ , \Pi _+ )} \, \big) \,
\big( \, \sqrt{g(\Pi _- , \Pi _- ) \, g(\Pi _+ , \Pi _+ )} -
g(\Pi _+ , \Pi _- ) \, \big) \, . 
\end{equation}
Clearly, (\ref{eq:eigen}) holds if and only if $\pi (n, \tau ) = 0$.
As $\Pi _+$ and $\Pi _-$ are linearly independent, the last bracket in 
(\ref{eq:proofpi}) is different from zero. So (\ref{eq:eigen}) is,
indeed, equivalent to (\ref{eq:defdis}).

A generic timelike surface $\Sigma$ is the worldsheet of a track that
appears curved to the eye of any observer, because $\Pi _+$ and $\Pi _-$
are non-zero. As (\ref{eq:gyro}) cannot be satisfied, a gyroscope
carried along the track cannot stay parallel to the track. With respect
to any observer on the track, Coriolis and centrifugal acceleration are 
linearly independent. The distinguished observer field is characterized 
by producing a symmetry between the backward and the forward directions.

Finally, we note that the vertex is not the only point on the 
characteristic hyperbola that is distinguished. As an alternative, 
we may consider the point which is closest to the origin, denoted by 2 
in Figure \ref{fig:hyper}. This gives us a second distinguished observer 
field. Physically, it is characterized by the fact that the total 
inertial acceleration perpendicular to the track becomes minimal. In 
contrast to the (first) distinguished observer field, the second 
distinguished observer field is not necessarily unique; there may 
be one or two such observer fields, corresponding
to the fact that a circle can be tangent to a hyperbola in one or 
two points. More importantly, the second distinguished observer
field is not invariant under conformal transformations; this
follows from our earlier observation that a conformal transformation
corresponds to a translation of the characteristic hyperbola. As in
this article we focus on conformally invariant properties, the
second distinguished observer field is of less interest to us.

\section{Special timelike surfaces of the first kind}\label{sec:first}
If $\Sigma$ is special of the first kind, the angle between the two
asymptotes in Figure~\ref{fig:hyper} is zero. Thus, the characteristic
hyperbola degenerates into a straight line which is run through twice,
with a turning point at the tip of the arrow (\ref{eq:turning}).
This turning point corresponds to the distinguished observer field
which is still determined by (\ref{eq:defdis}). However, now it
satisfies even the stronger condition $\Pi _+ = \Pi _-$.
The distinguished observer field is no longer characterized by the 
eigenvalue equations (\ref{eq:eigen}) because these equations are
now satisfied by \emph{any} orthonormal basis $(n , \tau )$. Note,
however, that now (and only in this case) the distinguished
observer field satisfies the ``strong'' eigenvalue
equation $\Pi (n, \, \cdot \, ) = \Lambda \otimes g(n, \, \cdot \, )$,
with an ``eigenvalue'' $\Lambda \, \in \, P^{\perp} (T \Sigma )$. 
 
If $\Sigma$ is special of the first kind everywhere, a track modeled
by $\Sigma$ appears curved to the eye of any observer, because $\Pi _+$
and $\Pi _-$ are non-zero. The distinguished observer field satisfies
(\ref{eq:gyro}) which means that a gyroscope carried by a distinguished
observer remains parallel to the track if it was so initially. For all 
other observers this is not true. If we write (\ref{eq:defgCc}) for 
the case that $n$ is the distinguished observer field, we read that
the Coriolis acceleration is zero for all $v$ whereas the centrifugal 
acceleration is non-zero for all $v \neq 0$.

\section{Special timelike surfaces of the second kind}\label{sec:second}
If $\Sigma$ is special of the second kind, the angle between the two
asymptotes in Figure \ref{fig:hyper} is $180^o$. Thus, the characteristic
hyperbola degenerates into a straight line which extends from infinity to 
infinity, passing through the tip of the arrow $\frac{1}{2} \Pi _0$.
This point corresponds to the distinguished observer field
which is still determined by (\ref{eq:defdis}). However,
now it satisfies the stronger condition $\Pi _+ = - \Pi _-$.
The distinguished observer field cannot be characterized by the 
eigenvalue equations (\ref{eq:eigen}) because the bilinear form 
$\pi$ is identically zero. 
 
If $\Sigma$ is special of the second kind everywhere, a track modeled
by $\Sigma$ appears curved to the eye of any observer on the track, 
as $\Pi _+$ and $\Pi _-$ are non-zero. Condition (\ref{eq:gyro}) 
cannot be satisfied, so a gyroscope does not remain 
parallel to the track if it was so initially, for any observer. If 
we write (\ref{eq:defgCc}) for the case that $n$ is the distinguished 
observer field, we read that the centrifugal acceleration is zero for 
all $v$ whereas the Coriolis acceleration is non-zero for all $v \neq 0$.
The vanishing of the centrifugal force is a measurable property by
which the distinguished observer field is uniquely determined. 

\section{Special timelike surfaces of the third kind}\label{sec:third}
Recall that $\Sigma$ is a one-way photon surface if it
is special of the third kind everywhere.
A one-way photon surface is a timelike surface $\Sigma$ that
is ruled by one family of lightlike geodesics, whereas the 
other family of lightlike curves in $\Sigma$ is non-geodesic.
This implies that, if $\Sigma$ is the worldsheet of a track,
the track visually appears straight in one direction but
curved in the other.  

For a one-way photon surface one of the two vectors that span the 
asymptotes in Figure \ref{fig:hyper} becomes zero. Thus, the
characteristic hyperbola degenerates into a straight line which is
run through once, beginning at infinity and then asymptotically 
approaching the tip of the arrow $\frac{1}{2} \Pi _0$. There
is no distinguished observer field because one side of equation 
(\ref{eq:defdis}) is nonzero and the other is zero for all 
orthonormal bases. By the same token, (\ref{eq:gyro}) is never 
satisfied because the vector on one side of this equation is always
zero whereas that on the other side is never zero. Hence,
it is impossible to carry a gyroscope along the track 
modelled by $\Sigma$ in such a way that its axis stays 
parallel to the track.

From (\ref{eq:defgCc}) we read that, on a one-way photon surface,
the Coriolis acceleration $a_{\text{Cor}}$ and the centrifugal 
acceleration $a_{\text{cent}}$ are always parallel. Also, we 
read that both $a_{\text{Cor}}$ and $a_{\text{cent}}$ are 
necessarily non-zero for $v \neq 0$. 

One-way photon surfaces can be easily constructed, on any Lorentzian 
manifold, in the following way. Choose at each 
point of a timelike curve a lightlike vector, smoothly depending
on the foot-point. With each of these lightlike vectors as the
initial condition, solve the geodesic equation. The resulting
lightlike geodesics generate a smooth timelike surface in the
neighborhood of the timelike curve. Generically, this is a 
one-way photon surface. (In special cases it may be a photon 
surface.)

\section{Special timelike surfaces of the fourth kind}\label{sec:fourth}
We now turn to photon surfaces, i.e. to timelike surfaces which are
everywhere special of the fourth kind. The worldline of a track is
a photon surface if and only if it is ruled by two families of
lightlike geodesics. As already outlined, this implies that the 
track appears straight to the eye of any observer on the track. 

For a photon surface the characteristic hyperbola degenerates into
a single point, situated at the tip of the arrow $\frac{1}{2} \Pi _0$.
This fact clearly indicates that all observer fields have equal 
rights, i.e., there is no distinguished observer field. 

If $\Sigma$ is special of the fourth kind at a point $p$, of the three 
components $\Pi _+$, $\Pi _-$ and $\Pi _0$ only the last one 
is different from zero at $p$. Thus, 
$
\Pi (u,w) = - \frac{1}{2} g(u.w) \Pi _0 \; 
$
at $p$, i.e., the second fundamental form $\Pi$ is a 
multiple of the first fundamental form $g$. Points 
where this happens are called \emph{umbilic points}.
A submanifold is called \emph{totally umbilic} if
all of its points are umbilic. Thus, a timelike surface
is a photon surface if and only if it is totally umbilic.
For a more detailed discussion of totally umbilic
submanifolds of semi-Riemannian manifolds see 
\cite{Perlick2005}.

The defining property $\Pi _+ = \Pi _- = 0$ of photon
surfaces implies that (\ref{eq:gyro}) is satisfied
for all orthonormal bases $(n, \tau )$. Hence, a
gyroscope that is initially tangent to the track
modelled by $\Sigma$ remains tangent to the track
forever, independent of its motion along the track.
This property characterizes photon
surfaces uniquely.

From (\ref{eq:defgCc}) we read that for a photon 
surface Coriolis acceleration $a_{\text{Cor}}$ and
centrifugal acceleration $a_{\text{cent}}$ vanish
identically. Again, this property characterizes 
photon surfaces uniquely. 

The most obvious example for a photon surface is a 
timelike plane in Minkowski spacetime. A less trivial 
example is the timelike surface $\vartheta = \pi /2$,
$r=3m$ in Schwarzschild spacetime. Inertial forces 
and gyroscopic transport on this circular track were 
discussed in several papers by Marek 
Abramowicz with various co-authors, see e.g. 
\cite{AbramowiczCarterLasota1988} and
\cite{Abramowicz1990}.

The existence of a photon surface requires a non-trivial 
integrability condition, so it is not guaranteed
in arbitrary Lorentzian manifolds, see 
\cite{FoertschHassePerlick2003}. In the same paper we
have given several methods of how to construct photon 
surfaces. Also, we have determined all photon surfaces
in conformally flat Lorentzian manifolds, and some 
examples of photon surfaces in Schwarzschild and 
Goedel spacetimes. 

\section{Example: Timelike surfaces in Kerr-Newman spacetime}\label{sec:example}
As an example, let $g$ be the Kerr-Newman metric in Boyer-Lindquist 
coordinates (see, e.g., Misner, Thorne and Wheeler 
\cite{MisnerThorneWheeler1973}, p.877)
\begin{equation}\label{eq:kerr}
  g =   - \frac{\Delta}{\rho ^2} \, \big( \, dt \, - \, 
  a \, \mathrm{sin} ^2 \vartheta \, d \varphi \big) ^2 \, + \,
  \frac{\mathrm{sin} ^2 \vartheta}{\rho ^2} \, \big(
  (r^2 + a^2) \, d \varphi \, - \, a \, dt \, \big) ^2 \, + \, 
  \frac{\rho ^2}{\Delta} \, dr^2 \,  + \, \rho ^2 \, d \vartheta ^2 \, ,
\end{equation}
  where $\rho$ and $\Delta$ are defined by
\begin{equation}\label{eq:rhodelta}
  \rho ^2 = r^2 + a^2 \, {\mathrm{cos}} ^2 \vartheta
  \quad \text{and} \quad 
  \Delta = r^2 - 2mr + a^2 + q^2 \, ,
\end{equation}
  and $m$, $q$ and $a$ are real constants.  We shall assume that 
\begin{equation}\label{eq:ma}
  0 \, < \, m \: , \quad 0 \, \le \, a \: , \quad \sqrt{a^2 + q ^2} \, \le \, m \, .
\end{equation}
In this case, the Kerr-Newman metric describes the spacetime around a rotating
black hole with mass $m$, charge $q$, and specific angular momentum $a$. The 
Kerr-Newman metric (\ref{eq:kerr}) contains the Kerr metric ($q=0$), the
Reissner-Nordstr{\"om} metric ($a=0$) and the Schwarzschild metric ($q=0$ and $a=0$)
as special cases which are all discussed, in great detail, in Chandrasekhar 
\cite{Chandrasekhar1983}.

By (\ref{eq:ma}), the equation $\Delta = 0$ has two real roots,
\begin{equation}\label{eq:hor}
r_{\pm} = m \pm \sqrt{ m^2 - a^2 - q ^2} \, ,
\end{equation}
which determine the two horizons. We shall restrict to the region 
\begin{equation}\label{eq:M+}
  M : \quad r_+ < r < \infty \, ,
\end{equation}
which is called the {\em domain of outer communication\/} of the Kerr-Newman 
black hole. 

For $0 < \vartheta < \pi$ and $r_+ < r < \infty$,
let $\Sigma _{\vartheta, r}$ denote the set of all points in $M$ where $\vartheta$
and $r$ take the respective values. Clearly, $\Sigma _{\vartheta , r}$ is a 
smooth two-dimensional timelike submanifold of $M$ homeomorphic to the 
cylinder $\mathbb{R} \times S^1$, parametrized by the cordinates $t$ and
$\varphi$. We may interpret $\Sigma _{\vartheta , r}$ as the worldsheet of
a circular track around the rotation axis of the black hole. We want to investigate 
for which values of $\vartheta$ and $r$
the timelike surface $\Sigma _{\vartheta, r}$ is special.
We choose the orthonormal basis
\begin{equation}\label{eq:E}
  n = \frac{1}{\rho \, \sqrt{\Delta}} \Big( (r^2 + a^2) \partial _t 
  + a \partial _{\varphi} \Big) \: , \qquad
  \tau = \frac{1}{\rho \, {\mathrm{sin}} \, \vartheta} \, 
  \big( \partial _{\varphi} + a \, {\mathrm{sin}} ^2 \vartheta \, \partial _t \big) \: .
\end{equation}
By a straight-forward calculation, we find the components $\Pi _{\pm}$
of the second fundamental form with respect to the lightlike basis 
$l_{\pm} = n \pm \tau$,
\begin{gather}\label{eq:PipmKN2}
\Pi _{\pm} \, = \, - \, \big( \, 2a^2 \, \text{sin}^2 \,  \vartheta
+ \rho^2 \pm 2a\sqrt{\Delta} \text{sin} \, \vartheta \, \big)
 \, 
\frac{\text{cos} \, \vartheta}{\rho^4 \text{sin} \, \vartheta} \, 
\partial _{\vartheta}
\\
\nonumber
\,  - \,
\big(  2r \Delta - (r-m) \rho ^2 \pm 2ar \sqrt{\Delta} \, \text{sin} \, 
\vartheta \big) \, \frac{1}{\rho^3 } \, \partial _r \; .
\end{gather}
$\Pi _+$ and $\Pi _-$ are linearly dependent if and only if $\vartheta = \pi /2$. Thus,
our circular track gives a special timelike surface if and only if it is in the equatorial plane.
The equation $\Pi _{\pm} =0$ is equivalent to $\vartheta = \pi /2$ and 
\begin{equation}\label{eq:crit}
  2 \, r \, \Delta - (r-m) \, r^2
  \pm 2 \, a \, r \, \sqrt{\Delta} \, =0 \; .
\end{equation}
For each sign, there is precisely one real solution $r_{\pm}^{\mathrm{ph}}$ to 
this equation in the domain of outer communication. Thus, among our timelike
surfaces $\Sigma _{\vartheta , r}$ there are precisely two one-way photon 
surfaces. They correspond to the well known co-rotating and counter-rotating
circular photon paths in the Kerr-Newman metric. In the Reissner-Nordstr{\"o}m
case, $a=0$, they coincide and give a photon surface at 
$r= \frac{3m}{2} +  \sqrt{\frac{9 m^2}{4} - 2 q^2}$ (cf., e.g., 
Chandrasekhar \cite{Chandrasekhar1983}, p.218).

For $\vartheta = \pi /2$ and $r > r_{+}^{\mathrm{ph}}$ or $r < r_{-}^{\mathrm{ph}}$,
the timelike surface $\Sigma _{\vartheta , r}$ is special of the first kind, for
$\vartheta = \pi /2$ and $r_{-}^{\mathrm{ph}} < r < r_{+}^{\mathrm{ph}}$ it is 
special of the second kind.

\bibliography{ref}

\end{document}